\documentclass[final]{aipproc}

\def\selectedlayoutstyle{8x11single}
\layoutstyle\selectedlayoutstyle

\SetInternalRegister\hbadness{8000} 

\newcommand\doingARLO[2][]{%
  \ifx\mmref\undefined #1\else #2\fi
}

\def\degr{\hbox{$^\circ$}}
\def\arcmin{\hbox{$^\prime$}}

\def\fdg{\hbox{$.\!\!^\circ$}}
\def\farcm{\hbox{$.\mkern-4mu^\prime$}}

\newcommand{\HI}{{\rm H\,\scriptstyle I}}
\newcommand{\HII}{{\rm H\,\scriptstyle II}}

\begin{document}

\title
      [Galactic Polarization]
      {Polarization Surveys of the Galaxy}


\author{Wolfgang Reich, Ernst F\"urst, Patricia Reich, Richard
        Wielebinski and \newline
        Maik Wolleben}{
  address={Max-Planck-Institut f\"ur Radioastronomie,
Auf dem H\"ugel 69, 53121 Bonn, Germany},
  email={wreich@mppifr-bonn.mpg.de},
  email={efuerst@mppifr-bonn.mpg.de},
  email={preich@mppifr-bonn.mpg.de},
  email={p022rwi@mppifr-bonn.mpg.de},
  email={wolleben@mppifr-bonn.mpg.de},
}

\copyrightyear  {2001}

\begin{abstract}
We report on sensitive $\lambda21$~cm and $\lambda11$~cm polarization
surveys of the Galactic plane carried out with the Effelsberg
100-m telescope at arcmin angular resolution and some related work.
Highly structured polarized emission is seen along the Galactic plane
as well as up to very high Galactic latitudes. These observations
reflect Faraday effects in the interstellar medium. Polarized
foreground and background components along the line of sight, modified
by Faraday rotation and depolarization, add in a complex way.
The amplitudes of polarized emission features are highly frequency dependent.
Small-scale components decrease in amplitude rapidly with increasing
frequency. We stress the need for sensitive absolutely calibrated
polarization data. These are essential for baseline setting and a correct
interpretation of small-scale structures. Absolutely calibrated data
are also needed to estimate the high-frequency polarized background. A
recent study of polarized emission observed across the local
Taurus-Auriga molecular cloud complexes indicates excessive synchrotron
emission within a few hundred parsecs. These results suggest that
possibly a large fraction of the Galactic high latitude total
intensity and  polarized emission is of local origin.
\end{abstract}

\date{\today}

\maketitle

\section{Survey History}

\ifthenelse{\equal\selectedlayoutstyle{6x9}}{\par\bfseries
  Note: The entire paper will be reduced 15\% in the printing
  process. Please make sure all figures as well as the text within the
  figures are large enough in the manuscript to be readable in the
  finished book.\par\bfseries
  Note: The entire paper will be reduced 15\% in the printing
  process. Please make sure all figures as well as the text within the
  figures are large enough in the manuscript to be readable in the
  finished book.\par\bfseries
  Note: The entire paper will be reduced 15\% in the printing
  process. Please make sure all figures as well as the text within the
  figures are large enough in the manuscript to be readable in the
  finished book.\normalfont}{}

Survey work has a tradition at the Max-Planck-Institut f\"ur
Radioastronomie in Bonn. At first there are the all-sky surveys at
408~MHz ($\lambda73$~cm, Haslam et al. {\cite{Haslam82}})
and the recently completed $\lambda21$~cm all-sky survey
(Reich {\cite{Reich82}}, Reich \& Reich {\cite{Reich86}} and Reich et
al. {\cite{Reich2001}}). Both surveys have a comparable angular resolution
of $0\fdg 8$ or $0\fdg 6$. Their sensitivities of 2~K or 0.05~K
match for spectral studies of the Galactic synchroton emission. Higher
angular resolution surveys at $\lambda21$~cm and $\lambda11$~cm
were carried out with the Effelsberg 100-m telescope, but had to be
limited to the Galactic plane because of the large amount of observing
time needed. These and other survey data including all references are
accessible via the internet at
\url{http://www.mpifr-bonn.mpg.de/survey.html}.

Although a number of new total intensity surveys were carried out in
the last two decades, polarization measurements were still just
available from a number of long wavelengths surveys, where
$\lambda21$~cm is the shortest wavelength. These surveys
were carried out in the sixties with the Dwingeloo 25-m telescope
(Brouw \& Spoelstra {\cite{Brouw76}}). Although these data have
rather moderate angular resolution they are quite carefully
absolutely calibrated. The Dwingeloo surveys cover large sections of
the northern sky, although they are not fully sampled. Large
fractional polarizations are noted at high Galactic latitudes and a
fairly smooth intensity and vector distribution is seen. Rotation
Measures (RMs) are small in general. These results did not immediately
trigger survey projects aiming for higher angular resolution.

The motivation for systematic polarization work at the Effelsberg
telescope is mainly based on the unexpected $\lambda11$~cm polarization
results of Junkes et al. {\cite{Junkes1987}}, which came out as a
byproduct of the first section on the Effelsberg $\lambda11$~cm
Galactic plane survey (Reich et al. {\cite{Reich1984}}). Beside
polarized sources, like supernova remnants, these data show
polarized emission patches with no apparent corresponding total
intensity structure. These have to be associated to the unresolved
diffuse background emission. It could be demonstrated by the
anticorrelation of diffuse thermal emission and integrated polarized
emission that a fraction of the emission must originate at a few kpc
distance in the Galactic disk (Junkes et al. {\cite{Junkes87b}}). This
indicates a highly structured interstellar medium, where "holes" with
low Faraday effects allow a study of the Galactic magnetic field at
large distances.

Sensitive high resolution polarization observations at $\lambda90$~cm
by Wieringa et al. {\cite{Wieringa93}} with the Westerbork telescope
showed filamentary features on degree scales at high latitudes. Their
origin has to be local. The total absence of enhanced synchrotron
emission calls for an explanation by a local Faraday screen modifying
the polarized background. However, a quantitative analysis has some
difficulties. The detected filaments at $\lambda90$~cm disappear at
shorter wavelengths, which again strengthens the case of a Faraday
screen. Because of the $\lambda^2$ dependence of the Faraday rotation
even small RMs have large effects at long wavelengths. We refer to
Sokoloff et al. {\cite{Sok98}} for a detailed discussion of Faraday
effects.

\begin{figure}
  \includegraphics[height=17cm,angle=270]{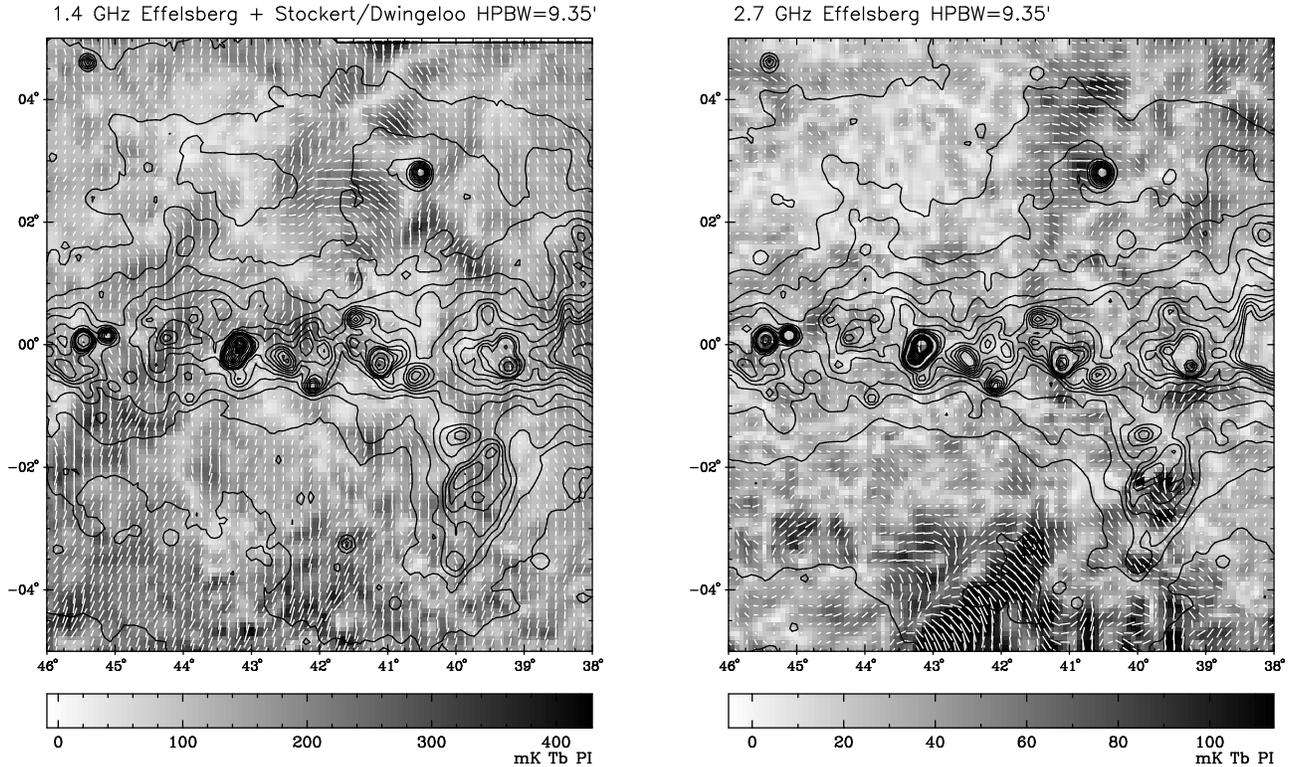}
 \caption {Section of the Galactic plane at $\lambda21$~cm and
$\lambda11$~cm shown at the same angular resolution.
The $\lambda21$~cm map was absolutely calibrated both for total
intensity and polarization using Stockert (Reich \& Reich
{\cite{Reich86}}) and Dwingeloo (Brouw \& Spoelstra {\cite{Brouw76}})
data, while the $\lambda11$~cm map is not on an absolute scale.
Polarized structures vary largely in intensity and angle.}
\end{figure}

\section{Galactic Plane Surveys including polarization}

From the Junkes et al. {\cite{Junkes1987}} $\lambda11$~cm
data it is obvious that low resolution data miss details of Galactic
polarized emission. It is also clear that the fine structure is quite
weak. The $\lambda11$~cm survey data need some smoothing to obtain a
sufficient signal-to-noise. Since the distribution of polarized
emission is not predictable from total intensity we started in 1994
the "$\lambda21$~cm Medium Galactic Latitude Survey" to map the
Galactic plane within $\pm 20\degr$ at $9\farcm 35$ (HPBW) with a
sensitivity equivalent to the total intensity confusion limit
of 7~mJy/beam or $\sim 15$~mK $\rm T_{b}$. This sensitivity was reached
with an integration time of 2~sec per $4\arcmin$ pixel. Thus the
surveyed area of $\rm \sim 7600\, deg^2$ needs $\sim 1000$h of
net integration time. Observations were exclusively done at night time
otherwise polarized solar emission shows up, which was picked up by far
sidelobes. Varying interference sometimes forces to stop
observations for several months. Nevertheless, in September 2001 the
survey is observed to $\sim 95$\%. Of course, the completion of data
reduction needs some more time. Basic data of the survey are listed in
Table~1.

\begin{table}
\begin{tabular}{lcc}
\hline
  \tablehead{1}{c}{b}{Frequency [GHz]}
  & \tablehead{1}{c}{b}{1.4}
  & \tablehead{1}{c}{b}{2.695} \\
\hline
HPBW & 9\farcm 35 & 5\farcm 1 \\
RMS PI [mK] & 8 & 11 \\
L-Coverage & $\sim 30\degr$ -- $220\degr$ & $4\fdg 9$ -- $76\degr$ \\
B-Coverage& +/-$20\degr$ & +/-$5\degr$\\
\hline
\end{tabular}
\caption{Basic observational parameters of the Effelsberg
$\lambda21$~cm and $\lambda11$~cm polarization surveys of the Galactic
plane}
\label{tab:a}
\end{table}

The observing and calibration methods of the "$\lambda21$~cm
Medium Galactic Latitude Survey" were already described in detail by
Uyan{\i}ker et al. {\cite{Bu98}}. These include some additional steps
to the standard Effelsberg reduction, in particular reduction
of spurious polarization from cross-talk and absolute calibration using
Dwingeloo survey data (Brouw \& Spoelstra {\cite{Brouw76}}), where
available. First maps of typical regions ranging from the first
quadrant towards the Galactic anticentre region are shown by
Uyan{\i}ker et al. {\cite{Bu99}}. The characteristics of the total
intensity are a smooth dominating background with faint ridges, arcs
and complex emission regions superimposed. Diffuse polarized emission
is seen everywhere. There are numerous polarized features without any
associated total intensity structures. In particular remarkable
depolarized loops, arcs and straight filaments are seen. These are most
pronounced towards the anticentre direction, where the absolute
intensities are low and the line of sight out of the Galaxy is short.
The standard interpretation is the assumption of a highly polarized
smooth background, which is seen through spatially varying Faraday
screens. We refer to Uyan{\i}ker et al. {\cite{Bu99}} for maps
illustrating these findings. A section of the Galactic plane is
shown in Fig.~1.

In addition the work of Junkes et al. {\cite{Junkes1987}} was continued
by analysing more polarization data of the Effelsberg $\lambda11$~cm
survey of the first quadrant, which  was extended from the initial
latitude coverage of $\pm 1\fdg 5$ to $\pm 5\degr$ (Reich et al.
{\cite{Reich1990}}). The results of the analysis of the polarization
data were published by Duncan et al. {\cite{Duncan99}}. Figure~1
shows some data from that work at the angular resolution of the
$\lambda21$~cm survey. Duncan et al. {\cite{Duncan99}} made also some
comparison with the Parkes $\lambda13$~cm polarization survey of the
fourth quadrant (Duncan et al. {\cite{Duncan97}}). The $\lambda11$~cm
maps show a clear increase of polarized emission with Galactic
latitude where depolarization is smaller. This result strengthens the
conclusion of Junkes et al. {\cite{Junkes87b}} on the kpc-origin of
some polarized emission. Duncan et al. {\cite{Duncan99}} noted an
anticorrelation of polarized emission at longitudes between
$20\degr$ and $45\degr$ with $\HI$ gas at kinematic distances
in the range 2~kpc~--~2.5~kpc, which requires the polarized emission to
originate at the same or larger distances.

\section{Some follow-up work}

Meanwhile some follow-up observations of $\lambda21$~cm and
$\lambda11$~cm polarization features have been started at
$\lambda6$~cm, although the required typical sensitivities of 1~mK and
below limit observations to a few fields only. Some first results have
been described by Reich et al. {\cite{Reich99}}. In brief: towards the
first Galactic quadrant the situation appears rather complex.
$\lambda21$~cm, $\lambda11$~cm (see Fig.~1), but also $\lambda6$~cm
polarization maps might differ largely in the sense that features
visible at one wavelength disappear at the other, while new features
show up. This indicates a superposition of numerous emission layers
along the line of sight across the Galaxy, which is quite plausible
when looking at the structured polarized emission towards
the anticentre at $\lambda21$~cm (Uyan{\i}ker et al. {\cite{Bu99}}),
where the line of sight is much shorter. For a decomposition of these
components observations at a denser sampling in wavelengths are
required, but these are not yet available. Of course, narrow-band
polarimetry within each available band will be quite helpful as well.

Towards the Galactic anticentre Reich et al. {\cite{Reich99} find
the situation to be less complex. For example: a ring-like structure of
$\sim 1\degr$ in diameter centered at $l,b = 192\fdg 2,\ 9\fdg 4$ was
also visible in polarization at $\lambda11$~cm and $\lambda6$~cm,
although it becomes very faint. A RM of $\rm \sim 120\, rad\, m^{-2}$
was derived. RM is calculated from: RM [$\rm rad\, m^{-2}$] =
0.81~$\rm n_{e}[cm^{-3}]~B_{||}[\mu G]~l[pc]$. Assuming distances
between 0.5~kpc and 3~kpc for this feature, its size l is between 9~pc
and 52~pc. Here a spherical shape is assumed. Upper limits for the
electron density are $\rm 1.5\, cm^{-3}$ or $\rm 0.6\, cm^{-3}$
from the emission measure of less than $\rm 20\, pc\, cm^{-6}$ set
by the noise of the $\lambda6$~cm total intensity signal. The
magnetic field component along the line of sight calculates lower
limits between $11\, \mu$G and $5\, \mu$G. It is not clear what
processes create such unusual magnetic-ionic structures high above the
Galactic plane. We note that for this and similar features the spectrum
of the polarized signal is steeper than the total intensity spectrum.

\subsection{The Taurus-Auriga region and the local synchrotron emissivity}

The $\lambda21$~cm Medium Galactic Latitude Survey data from
longitudes $150\degr$ to $190\degr$ and latitudes from $-4\degr$ to
$-20\degr$ have recently been reduced and analysed by Wolleben
{\cite{Wolleben2001}}. This region includes the well studied
Taurus-Auriga and Perseus molecular cloud complexes, which are located
at distances of $\sim 140$~pc and 350~pc. The molecular clouds are
partly seen in superposition.  CO survey data and IRAS data show highly
structured emission across this area. Molecular gas and dust clouds
are partly well correlated. There is an enhancement of polarized
emission in the direction of these molecular cloud complexes. However,
the total intensity smoothly increasing towards the Galactic plane but
there is no indication of enhanced synchrotron emission. A number of
discrete polarized features ranging between $\sim 0.3\degr$ and
$1\degr$ in diameter could be identified adjacent to molecular and dust
clouds. This association suggests the existence of a Faraday screen at
the same distance. Polarized emission originating behind the molecular
material gets modified and adds with the foreground polarization in a
different way than outside the Faraday screen. Wolleben
{\cite{Wolleben2001}} noticed for a number of polarized structures a
clear systematic dependence of the polarization angle with polarized
intensity. Nine objects including their surroundings were modelled
assuming regular foreground and background polarization. The
background emission is then subject to Faraday modulation towards the
polarized feature. Pure Faraday rotation as well as Faraday rotation
with correlated depolarization were considered and fitted to the
polarization angle--polarized intensity relation. RMs of up to
$\rm \sim 30\, rad\, m^{-2}$ were derived, quite much for clouds of
a maximum extent between 0.7~pc and 2.4~pc (distance 140~pc) or
1.8~pc to 6~pc (distance 350~pc).  For a number of polarized
features both models fit the data quite well and a second frequency is
needed for a distinction between both models. The average of
$\lambda21$~cm polarized intensities in front and behind the nine the
Faraday screens near the molecular clouds have values of $\sim 220$~mK
at polarization angle $25\degr$ and 290~mK at $-35\degr$ for the
foreground and background components, respectively. Both components are
related to a regular magnetic field. Both values are uncertain by
$\sim 100$~mK and $\sim 20\degr$, respectively. Interestingly, a
rather similar jump in polarization angle is seen from stellar
polarization data (Heiles {\cite{Heiles2000}}) in the range up to
200~pc and 200~pc to 500~pc. The E-vectors agree on average within
$20\degr$ with those from the radio data, where the internal scatter
is about the same.

Polarized intensity  of $\sim 0.5$~K from a regular magnetic field
originating within 0.5~kpc requires a total intensity synchrotron component
of at least 0.7~K. This is a lower limit from the intrinsically
polarized regular field. Estimates for the fraction of the regular
magnetic field component range from 60\% to 70\% of the total field (see
Beck {\cite{Beck2000}} for a review and references). If these estimates
also hold for the very local magnetic field the $\lambda21$~cm
synchrotron component raises to $\sim 1$~K within 0.5~kpc.

Beuermann et al. {\cite{Beuer85}} have unfolded the $\lambda73$~cm
all-sky survey (Haslam et al. {\cite{Haslam82}) for a three-dimensional
radio emission model of the Galaxy. They quote a local synchrotron
emissivity of $\sim 11.7$~K/kpc, which they found to be in
agreement with previous studies. This scales to $\sim 0.3$~K/kpc at
$\lambda21$~cm, which is significantly below the 1.4~K/kpc to 2~K/kpc
for the local emissivity towards the Taurus-Auriga complex. In this
direction the total Galactic emission above the cosmic microwave
background is at the 2~K level. However, this very local synchrotron
component depends strongly on the magnetic field direction and
therefore is unlikely isotropic. It is interesting to note that the
recently derived average local emissivity at 22~MHz by Roger et al.
{\cite{Roger99}} is about three times higher than the value adopted by
Beuermann et al. {\cite{Beuer85}}. The Roger et al. {\cite{Roger99}}
result is based on emission towards extended opaque $\HII$ regions at
distances between 400~pc and 2.2~kpc.

Enhanced local emissivity implies that emission at high Galactic
latitudes is more affected by small scale fluctuations in the
interstellar medium. This is expected to be the case in particular for
polarized emission. The thick disk shrinks in size and its smooth
intensity distribution is less dominant. Cosmic microwave studies at
high Galactic latitudes should take this result into account.

\section{Absolute Calibration}

\begin{figure}
  \includegraphics[height=17cm,angle=270]{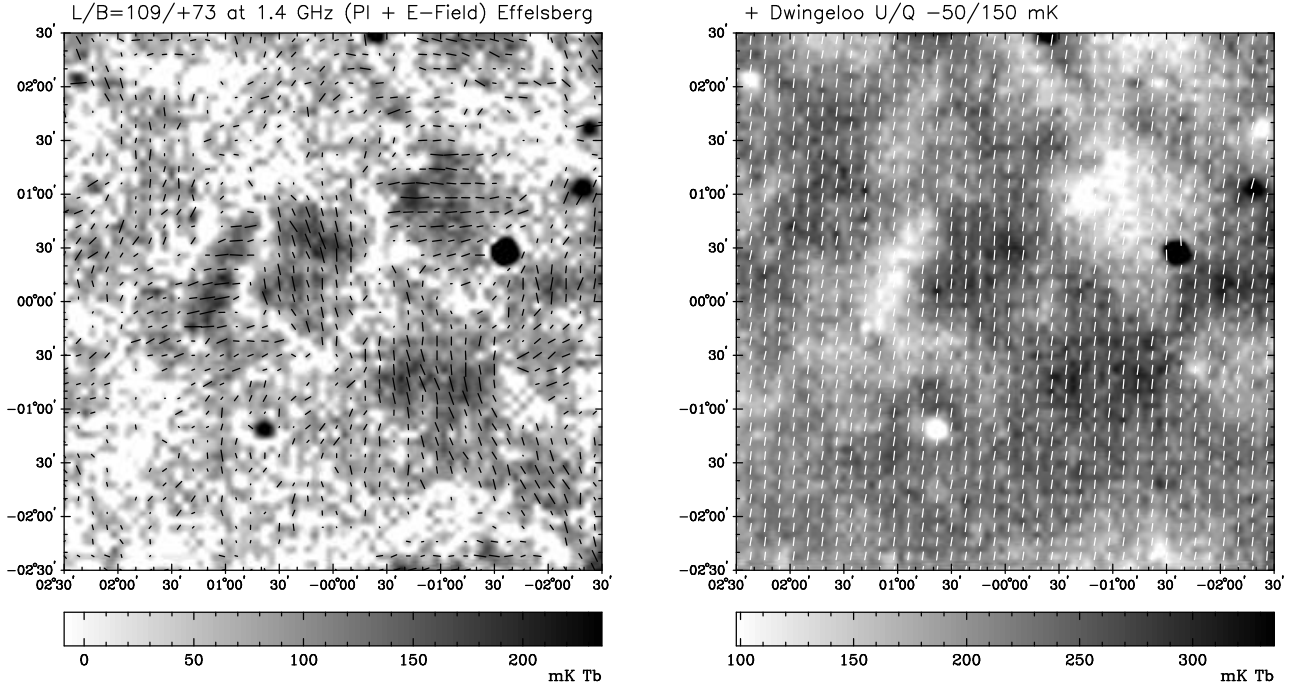}
 \caption{High latitude polarized emission at 1.4~GHz without (left)
and with (right) large-scale structure added. Low-resolution absolutely
calibrated Dwingeloo data were used to estimate the missing large-scale
componets of the Effelsberg data. Due to severe undersampling of the
Dwingeloo data just constant offsets for U ($-$50~mK) and Q (+150~mK)
have been added to the Effelsberg data. Depending on their polarization
angle relative to the large-scale emission small-scale structures
appear as enhancements or depressions superposed on the large-scale
structure. A number of polarized sources are visible. The total
intensity (not shown here) shows many compact sources in the field but
no extended structures possibly related to the polarized features.}
\end{figure}

The Effelsberg survey maps have a relative offset resulting from the
assumed "zero" at the edges of each individual observed field. Each
field is observed twice in Galactic longitude and latitude direction.
Just the strong emission of the Galactic ridge in the first quadrant
requires to scan along latitude direction only. The fields at
$\lambda21$~cm and $\lambda11$~cm  have typical sizes
of $\sim 10\degr \times 10\degr$ or $\sim 3\degr \times 3\degr$,
respectively. The combined observations have an "average zero" at their
boundaries. As described by Uyan{\i}ker et al. {\cite{Bu98} the total
intensity and polarization data were absolutely calibrated by available
lower angular resolution data from the Stockert and the Dwingeloo 25-m
telescopes. While the total $\lambda21$~cm intensities could be
completely corrected, the polarization data could not because of
incomplete mapping, undersampling and low sensitivity. The correction
effects at low latitude are largest for total intensities and low for
polarization, while at high latitudes the situation is reverse.
This is problematic, because the correct zero level in polarization is
much less predictable than for total intensities. Stokes U and Q may
have positive or negative values depending on the polarization angle
$\phi = 0.5$~atan(U/Q) and the polarized intensity calculates from
$\rm (U^{2}+Q^2)^{0.5}$. Adding large-scale dominating U and Q
components with the same or the opposite sign converts a small-scale
polarized emission feature into enhanced or reduced emission within the
large-scale structure. Figure~2 shows an example of this effect: A high
Galactic latitude $\lambda21$~cm Effelsberg map with relative zero
levels and with the large-scale components added. The large-scale
structure entirely dominates. The average polarized emission from the
Effelsberg map is about 8~mK, which is just 5\% of the large-scale
emission. However, the large-scale structure is estimated from a few
data points near the observed field. These U and Q values were averaged
and used to adopt the level of the Effelsberg U and Q maps. The
uncertainty of the large-scale amplitudes is at least 30\%. In
addition, possible gradients in U and Q or fluctuations on scales of a
degree or larger could not be taken into account.

The $\lambda11$~cm total intensity data could be corrected for
missing large-scale emission ($\ge 10\degr$), while the polarization
data from all fields were adjusted relative to each other. In fact it
was shown by  Duncan et al. {\cite{Duncan99}} that the $\lambda13$~cm
Parkes data contain emission from larger scales, because the survey was
combined from larger individual fields. The only indirect method to
adopt large-scale polarized structure at shorter wavelengths
than $\lambda21$~cm is to extrapolate by assuming a spectral index for
the polarized emission and to adopt a long wavelength RM in part
available from the work of Brouw \& Spoelstra {\cite{Brouw76}}. New
absolutely calibrated data at $\lambda21$~cm at mK-sensitivity are
needed, which may use the already existing Dwingeloo data for
adjustment purpose, as was discussed by Reich \& Wielebinski
{\cite{Reich01}}. Also plans for sensitive $\lambda6$~cm observtions
exist, where the angular resolution of a 25-m telescope matches that of
the Effelsberg $\lambda21$~cm observations. At $\lambda6$~cm
and shorter wavelengths extrapolation from the Brouw \& Spoelstra
{\cite{Brouw76}} data seem irrelevant.

\section{Concluding remarks}
We have discussed some work on polarized Galactic emission from the
Effelsberg $\lambda21$~cm and $\lambda11$~cm surveys including
some follow-up observations and analysis. Small-scale Faraday
modulation causes highly fluctuating polarized or depolarized
structures. The origin of the corresponding magneto-ionic
structures -- enhancement of thermal electron density or compression of
the magnetic field -- remains open so far. Small-scale emission becomes
rather faint at shorter wavelengths. The dominating polarized
background appears to be smooth on degree scales and larger. It is
likely the dominating component at short wavelength. Its investigation
requires very sensitive absolutely calibrated measurements. This needs
substantial efforts and much telescope time, but can be done with small
(25-m class) telescopes. At high latitudes the contribution from
polarized synchrotron emission originating within a few hundred parsecs
appears to be larger than previously assumed and fluctuations in
polarization are also visible at high latitudes not much different from
those seen at low latitudes.

\begin{theacknowledgments}
We like to thank A.~R. Duncan, N. Junkes and B. Uyan{\i}ker for their
contributions to the polarization survey work at Bonn.
\end{theacknowledgments}



\begin{thebibliography}{}
\bibitem{Haslam82}
 Haslam, C.~G.~T, Salter, C.~J., Stoffel, H., and Wilson, W.~E.,
 \emph{Astron. Astrophys. Suppl.}, {\bf 48}, 219 (1982).
\bibitem{Reich82}
 Reich, W., \emph{Astron. Astrophys. Suppl.}, {\bf 48}, 219 (1982).
\bibitem{Reich86}
 Reich, P., and Reich, W., \emph{Astron. Astrophys. Suppl.}, {\bf 48},
 219 (1986).
\bibitem{Reich2001}
 Reich, P., Testori, J.~C., and Reich, W., \emph{Astron. Astrophys.},
 {\bf 376}, 861 (2001).
\bibitem{Brouw76}
 Brouw, W.~N., and Spoelstra, T.~A.~Th., \emph{Astron. Astrophys.
 Suppl.}, {\bf 26}, 129 (1976).
\bibitem{Junkes1987}
 Junkes, N., F\"urst, E., and Reich, W., \emph{Astron. Astrophys.
 Suppl.}, {\bf 69}, 451 (1987).
\bibitem{Reich1984}
 Reich, W., F\"urst, E., Steffen, P., Reif, K., and Haslam, C.~G.~T.,
 \emph{Astron. Astrophys. Suppl.}, {\bf 58}, 197 (1984).
\bibitem{Junkes87b}
 Junkes, N., F\"urst, E., and Reich, W., in \emph{Interstellar Magnetic Fields},
 edited by R. Beck and R. Gr\"ave, Springer, Berlin, 1987, p. 115.
\bibitem{Wieringa93}
 Wieringa, M.~H., de Bruyn, A.~G., Jansen. D., Brouw, W.~N., and
 Katgert, P., \emph{Astron. Astrophys.}, {\bf 268}, 215 (1993).
 \bibitem{Sok98}
 Sokoloff, D.~D., Bykov, A.~A., Shukurov, A., Berkhuijsen, E.~M., Beck,
 R., and Poezd, A.~D., \emph{MNRAS}, {\bf 299}, 189 (1998).
\bibitem{Bu98}
 Uyan{\i}ker, B., F\"urst, E., Reich, W., Reich, P., and Wielebinski,
 R., \emph{Astron. Astrophys. Suppl.}, {\bf 132}, 401 (1998).
\bibitem{Bu99}
 Uyan{\i}ker, B., F\"urst, E., Reich, W., Reich, P., and Wielebinski,
 R., \emph{Astron. Astrophys. Suppl.}, {\bf 138}, 31 (1999).
\bibitem{Reich1990}
 Reich, W., F\"urst, E., Reich, P., and Reif, K., \emph{Astron.
 Astrophys. Suppl.}, {\bf 85}, 633 (1990)
\bibitem{Duncan99}
 Duncan, A.~R., Reich, P., Reich, W., and F\"urst, E., \emph{Astron.
 Astrophys.}, {\bf 350}, 447 (1999).
\bibitem{Duncan97}
 Duncan, A.~R., Haynes, R.~F., Jones, K.~L., and Stewart, R.~T.,
 \emph{MNRAS}, {\bf 291}, 279 (1997).
\bibitem{Reich99}
 Reich, W., Uyan{\i}ker, B., F\"urst, E., Reich, P., and Wielebinski,
 R., in \emph{Galactic Foreground Polarization}, edited by E.~M.
 Berkhuijsen, Workshop Proceedings, MPIfR Bonn, 1999, p. 54.
\bibitem{Wolleben2001}
 Wolleben, M., Diploma Thesis, Bonn University (2001).
\bibitem{Heiles2000}
 Heiles, C., \emph{Astron. J.}, {\bf 119}, 923 (2000).
\bibitem{Beck2000}
 Beck, R., in \emph{The Astrophysics of Galactic Cosmic Rays},
 edited by R. Diehl et al., Space Sience Review, Kluwer Academic
 Publishers, Dordrecht, 2001, in press.
\bibitem{Beuer85}
 Beuermann, K., Kanbach, G., and Berkhuijsen, E.~M., \emph{Astron.
 Astrophys.}, {\bf 153}, 17 (1985).
\bibitem{Roger99}
 Roger, R., Costain, C.~H., Landecker, T.~L., and Swerdlyk, C.~M.,
 \emph{Astron. Astrophys. Suppl.}, {\bf 137}, 7 (1999).
\bibitem{Reich01}
 Reich, W., and Wielebinski, R., in \emph{"Radio Polarization: A New
 Probe of the Galaxy"}, edited by T.~L. Landecker, Workshop Proceedings,
 DRAO, Penticton, 2001, p. 55.
\end{thebibliography}

{}

\end{document}